Comment: LaTeX file
-------------------------------------------
\documentstyle[preprint,aps]{revtex}
\begin{document}
\draft
\sloppy
\scrollmode

\title{\vspace{-5.5cm} ~ \hfill {\normalsize FUB-HEP/95}\\[4.5cm]
Path Integral of Relativistic Coulomb System}
\author{
H.~Kleinert\thanks{
 E-MAIL: kleinert@einstein.physik.fu-berlin.de; URL:
http://www.physik.fu-berlin.de/\~{}kleinert
}}
\address{Institut f\"{u}r Theoretische Physik,\\
          Freie Universit\"{a}t Berlin\\
          Arnimallee 14, D - 14195 Berlin}
\date{\today}
\maketitle
\begin{abstract}
The path integral
of the relativistic Coulomb system
is solved, and
the wave functions
are extracted from
the resulting amplitude.

\end{abstract}
\newpage
\noindent
1)~While the path integral of the nonrelativistic Coulomb system
has been solved some 15 years ago \cite{DK} and further
discussed by many authors \cite{rd} --\cite{CastrigianoS91},
so that it has become
textbook material \cite{PI},
the relativistic problem has remained open.
The purpose of this note is to fill this gap.

\noindent
2)~Consider first
 a relativistic free particle of mass $M$.
If $x( \lambda )$ describes its orbit in $D$ spacetime dimensions
in terms of some parameter $ \lambda $,
the
classical action reads
\begin{equation} \label{19.247}
{\cal A} _{\rm cl}= Mc \int_{\lambda _a}^{\lambda _b}
 d\lambda  \sqrt{  x'{} ^2 (\lambda )},
\end{equation}
where $c$ is the light velocity.
This action cannot be used to set up
a path integral
for the time evolution amplitude since it would not yield the
well-known Green function of a Klein-Gordon field.
An action which serves this purpose
can be constructed
with the help of an
auxiliary fluctuating variable $ \rho ( \lambda )$
and reads \cite{fh}
\begin{eqnarray} \label{19.245}
{\cal A} = \int_{\lambda  _a}^{\lambda  _b} d\lambda
\left[     \frac{M}{2\rho (\lambda  )}  x'{} ^2 (\lambda  )
    + \frac{Mc^2 }{2}\rho (\lambda  )\right] .
\end{eqnarray}
Classically, this action coincides with the original
${\cal A}_{\rm cl}$
since it is extremal for
\begin{equation} \label{19.246}
\rho (\lambda ) =  \sqrt{ x'{}^2 (\lambda )} /c.
\end{equation}
Inserting this back into (\ref{19.245})
we see that
${\cal A}$ reduces to
${\cal A}_{\rm cl}$.

The action ${\cal A} _{\rm cl}$ is
invariant under
arbitrary reparametrizations
\begin{eqnarray} \label{19.257}
\lambda  & \rightarrow  & f(\lambda ).
\end{eqnarray}
The action ${\cal A} $
shares this invariance,
if $ \rho ( \lambda )$ is simultaneously  transformed as
\begin{equation} \label{19.258}
\rho  \rightarrow  \rho / f'.
\end{equation}

The  action  ${\cal A} $
 has the advantage of being
quadratic in the orbital variable
$x( \lambda )$.   If the physical time
is analytically continued to imaginary values so that the metric
becomes euclidean,
the action looks like that of a
nonrelativistic particle moving as a function of a pseudotime $ \lambda $
through a $D$-dimensional
euclidean spacetime, with a mass depending on $ \lambda $.

\noindent
3)~
To set up a path integral, the action has to be pseudotime-sliced,
say at $ \lambda _0= \lambda _a, \lambda _1,\dots, \lambda _{N+1}= \lambda _b$.
 If $ \epsilon _n= \lambda _n- \lambda _{n-1}$ denotes the
thickness of the $n$th slice, the sliced
action reads

\begin{eqnarray} \label{19.252}
{\cal A}^N = \sum_{n=1}^{N+1} \left[ \frac{M}{2\rho _n\epsilon _n}
(\Delta  x _n)^2
 + \frac{Mc^2}{2}\epsilon _n\rho _n
          \right],
\end{eqnarray}
where  $ \rho _n\equiv  \rho ( \lambda _n)$,
 $ \Delta x_n\equiv x_n-x_{n-1}$, and
$  x_n\equiv x( \lambda _n)$.
A path integral
$\int ({\cal D}^Dx/ \sqrt{ \rho }^D)e^{-{\cal A}/\hbar }$ may be defined as the
limit
$N \rightarrow \infty$ of the product of integrals
\begin{eqnarray} \label{19.251}
\frac{1}{\sqrt{ 2\pi \hbar \epsilon _b\rho _b/M}^D}
\prod _{n=1}^{N} \left[\int  \frac{d^Dx_n}{\sqrt{ 2\pi \hbar
       \epsilon _n \rho _n/M}^D}\right]
       \exp \left( -\frac{1}{\hbar } {\cal A}^N\right) ,
\end{eqnarray}
This can immediately be evaluated, yielding
\begin{eqnarray} \label{19.253}
\frac{1}{\sqrt{ 2\pi \hbar L/Mc}^D} \exp \left[ -\frac{Mc}{2\hbar }
     \frac{( x _b- x _a)^2}{L}
     - \frac{Mc}{2\hbar }L \right] ,
\end{eqnarray}
where the quantity
\begin{equation} \label{19.254}
\!\!\!\!\!L\equiv c \sum_{n=1}^{N+1} \epsilon _n \rho _n
\end{equation}
has the continuum limit
\begin{equation} \label{19.255}
L =c \int_{\lambda _a}^{\lambda _b} d\lambda  \rho (\lambda ).
\end{equation}
Classically, this is the
 reparametrization invariant
 length of a path,
as is obvious after inserting (\ref{19.246}).

If the amplitude  (\ref{19.253}) is multiplied by $ \lambda _C/2$,
where $ \lambda _C=\hbar /Mc$ is the Compton wavelength of the particle,
an integral over all positive $L$
yields the correct Klein-Gordon amplitude
\begin{eqnarray}
   \left( { x}_b  \vert { x}_a  \right)
           =\frac{1}{(2\pi )^{D/2}}\left(  \frac{Mc}{ \hbar \sqrt{ x^2}}\right)
^{D/2-1} K_{D/2-1}
\left( {Mc} \sqrt{ x^2}/\hbar \right) ,
\label{19.5}\end{eqnarray}
where $K_\nu(z) $ is
the modified Bessel function.

The result does not depend on the choice of $ \rho ( \lambda )$, this being
 a manifestation of the reparametrization invariance.
We may therefore write the continuum version of the path integral
for the relativistic free particle
as
\begin{eqnarray}
     \left( { x}_b\vert { x}_a \right) =
\frac{1}{2}{\lambda_{\rm C}} \int^{\infty} _{0}
    dL \int {\cal D} \rho \Phi [\rho ]
\int \frac{{\cal D}^D x}{ \sqrt{ \rho }^D}  e^{-{\cal A}},
\label{19.1a}\end{eqnarray}
where $\Phi [\rho ]$ denotes
a convenient {gauge-fixing functional}, for instance
$\Phi [\rho ]=\delta [\rho -1]$ which fixes
$\rho (\lambda )$
 to unity everywhere.

To understand the factor $1/ \sqrt{ \rho }^D$ in the measure
of
(\ref{19.1a}),
we make use of the
canonical form of
the action (\ref{19.245}),
\begin{eqnarray} \label{19.248}
{\cal A}[p,x] = \int_{\lambda _a}^{\lambda _b}d\lambda  \left[  -i{ p}  x'
        +
        \frac{\rho (\lambda )p^2}{2M} +
        \frac{Mc^2}{2}  \rho (\lambda )\right] .
\end{eqnarray}
After pseudotime slicing, it reads
\begin{eqnarray} \label{19.249}
{\cal A}^N[p,x]  = \sum_{n=1}^{N+1}
    \left[ -ip_n ( x _n- x _{n-1})
 + \rho _n \epsilon _n \frac{p_n^2}{2M}
        + \frac{Mc^2 }{2}\epsilon _n\rho _n\right].
\end{eqnarray}
At a  fixed $ \rho (\lambda )$,
the path integral has then the usual canonical measure:
\begin{eqnarray} \label{19.250}
\int {\cal D}^Dx \int \frac{{\cal D}^Dp}{(2\pi \hbar)^D }
e^{-{\cal A}[p,x]/\hbar}
\approx
\prod _{n=1}^{N} \left[ \int d^Dx_n\right]
    \prod _{n=1}^{N+1} \left[ \int \frac{d^Dp_n}{(2\pi  \hbar )^D}
    \right] e^{-{\cal A}^N[p,x]/\hbar}.
\end{eqnarray}
By integrating out the momenta,
we obtain  (\ref{19.251})
with the action (\ref{19.252}).
%

\noindent
4)~
The fixed-energy amplitude is related to (\ref{19.1a})
by a Laplace transformation:
\begin{eqnarray}
   \left( {\bf x}_b \vert {\bf x}_a \right) _E \equiv
-i     \int^{\infty} _{x^0_a} dx^0_b e^{E(x^0  _b-x^0  _a)/\hbar }
\left( { x}_b  \vert
     { x}_a\right) ,
\label{19.int}\end{eqnarray}
where ${ x^0}$ denotes the temporal component
and
${\bf x}$ the purely spatial part of the $D$-dimensional vector $x$.
The poles and cut of $
 \left( {\bf x}_b \vert {\bf x}_a \right) _E
$ along the
energy axis
contain all information on the bound and continuous
eigenstates of the system.
The fixed-energy amplitude has  the path integral
representation
\begin{eqnarray}
   \left( {\bf x}_b \vert {\bf x}_a \right) _E =
 \int^{\infty} _{0}
    dL \int {\cal D} \rho \Phi [\rho ]
\int \frac{{\cal D}^{D-1} x}{ \sqrt{ \rho }^{D-1}}  e^{-{\cal A}_E/\hbar },
\label{19.int2}\end{eqnarray}
with the action
\begin{eqnarray}
   {\cal A} _E= \int^{\lambda _b}_{\lambda _a} d\lambda  \left[
     \frac{M}{2\rho (\lambda )} {\bf  x}'{}^2 (\lambda )
   -
\rho (\lambda )    \frac{E ^2}{2Mc^2}
  + \frac{Mc^2}{2} \rho (\lambda )\right].
\label{19.2p}\end{eqnarray}
This is seen by
writing the temporal part of the sliced $D$-dimensional
action (\ref{19.252})
in the canonical form (\ref{19.249}).
By integrating out
the temporal coordinates
$x^0_n$ in (\ref{19.250}), we obtain $N$ $ \delta $-functions.
These remove $N$ integrals over
the momentum
variables
$p^0_n$, leaving only a single integral over a common $p^0$.
The Laplace transform
(\ref{19.int}), finally, eliminates also this integral,
making $p^4$ equal to
$-iE/c$.
In the continuum limit,
we thus obtain the action (\ref{19.2p}).

\noindent
5)~
The path integral
(\ref{19.int2}) forms the
basis for studying relativistic particles in an
external time-independent potential $V({\bf x})$.
This
is introduced into the path integral
(\ref{19.int2})
by simply substituting the energy $E$ by $E-V({\bf x})$.

For an attractive  Coulomb potential in $D-1=3$ spatial dimensions,
the above substitution
changes the second term in the action
(\ref{19.2p})
to
\begin{eqnarray}
  {\cal A}_{\rm int} = - \int^{\lambda _b}_{\lambda _a} d\lambda
  \, \rho (\lambda ) \frac{( E + {e^2}/{|{\bf x}|}) ^2 }
{2Mc^2}
{}.
\label{19.8}\end{eqnarray}
The associated path integral is calculated
with the help of a
Duru-Kleinert transformation \cite{DK} as follows:

First, we increase
the three-dimensional configuration space
in a trivial way by a dummy forth component
$x^4$
(as in the nonrelativistic case).
The additional variable $x^4$ is eliminated at the end by
an integral $\int {d x_a^4}/{|{\bf x}_a|}=\int d\gamma _a$ (see
Eqs.~(13.114) and (13.121) in Ref. \cite{PI}).
Then we perform a
Kustaanheimo-Stiefel transformation
$ dx^\mu  = 2 A (u)^\mu {}_\nu du^\nu  $ (Eq.~(13.100) in Ref. \cite{PI}).
This changes
$x'{}^\mu {}^2$ into
$4 \vec u^2 \vec u'{\,}^2$,
with the vector symbol indicating the four-vector nature of
$\vec u$.
The transformed action reads
\begin{eqnarray}
&&  \! \hspace{-2cm}~ \,\tilde {\cal A}_E\! =\!
          \int^{\lambda _b}_{\lambda _a}\!d\lambda
           \bigg\{ \frac{4M\vec{u}^2}{2\rho (\lambda )}
              \vec{u}'{\,}^2(\lambda )
\!       + \!\frac{\rho (\lambda )}{2Mc^2 \vec{u}^2}
             \!  \left[ (M^2 c^4\! -E^2 )\vec{u}^2
     \!        - \!2Ee^2 \!-\! \frac{e^4}{\vec{u}^2}\right] \!\!\bigg\}.
\label{19.10}\end{eqnarray}
We now choose the gauge
$\rho (\lambda ) = 1$,
and go from $\lambda $
to a new parameter $s$
via the time transformation
$d\lambda  = fds$ with
$f = \vec{u}^2$. This leads to the Duru-Kleinert-transformed
action
\begin{eqnarray}
  &&\hspace{-2cm}{\cal A}_E^{\rm DK} =            \!
     \int^{s _b}_{s_a} ds \bigg\{ \frac{4M}{2}
           \vec  u'{\,}^2(s)
+ \frac{1}{2Mc^2}
 \left[ \left( M^2 c^4 - E^2\right) \vec u^2
            -2 Ee^2 - \frac{e^4}{\vec u^2}\right] \!\!\bigg\} .
\label{19.11}\end{eqnarray}
It describes a particle
of mass $\mu =4M$
moving as a  function of the ``pseudotime"
$s$
in a harmonic oscillator potential of frequency
\begin{eqnarray}
  \omega = \frac{1}{2M  c} \sqrt{  M^2 c^4 - E^2 }.
\label{19.12}\end{eqnarray}
The oscillator possesses an additional
attractive potential
$ -{e^4}/{2Mc^2} {\vec u^2} $ which is conveniently parametrized
in the form of a centrifugal barrier
\begin{eqnarray}
   V_{\rm extra}  =
  \hbar ^2 \frac{l_{\rm extra}^2}{2\mu \vec u^2},
\label{19.13a}\end{eqnarray}
whose  squared  angular momentum
has the negative value
\begin{eqnarray}
 l^2_{\rm extra} \equiv - 4\alpha ^2,
\label{19.13}\end{eqnarray}
where $\alpha $ denotes the feinstructure constant $\alpha \equiv e^2/\hbar
c\approx 1/137$.
In addition,
there is also a trivial constant potential
\begin{eqnarray}
 V_{\rm const} = - \frac{E}{Mc^2}e^2.
\label{19.14}\end{eqnarray}
If we ignore, for a moment, the
centrifugal barrier $V_{\rm extra}$,
the solution of the path integral can immediately be
written down
(compare Eq.~(13.121) in Ref. \cite{PI})
\begin{eqnarray}
 \!\!\!\!\!\!\!\! \left( {\bf x}_b \vert {\bf x}_a\right) _E
         = -i\frac{\hbar}{2Mc} \frac{1}{16}
         \int^{\infty} _{0} dL
        ~ e^{e^2{EL}/{Mc^2\hbar } }
         \int^{4\pi }_{0} {d \gamma _a} \left( \vec{u}_b L
          \vert \vec{u}_a 0 \right) ,
\label{19.16}\end{eqnarray}
where
$ \left( \vec{u}_b L \vert \vec{u}_a 0\right)$
is the time evolution amplitude
of the four-dimensional
harmonic oscillator.

There are no time-slicing corrections for the same reason as in the
three-dimensional
case.
This is ensured by the
affine connection of the
{Kustaanheimo-Stiefel transformation}
satisfying
\begin{equation} \label{19.pr}
\Gamma _{\mu }{}^{\mu \lambda  }=g^{\mu \nu }e_i{}^{\lambda} \partial _\mu
e^i{}_\nu =0
\end{equation}
(see the discussion in Section 13.6 of Ref. \cite{PI}).

A $ \gamma _a $-integration leads to
\begin{eqnarray} \label{19.13.n1}
({\bf x} _b|{\bf x}_a)_E & = & -i \frac{\hbar }{2Mc}\frac{M\kappa }{\pi \hbar}
     \int _{0}^{1} d\varrho  \frac{\varrho ^{-\nu }}{(1-\varrho )^2}
     I_0 \left( 2\kappa  \frac{2\sqrt{ \varrho }}
     {1-\varrho } \sqrt{( r_br_a +  {\bf x}_b{\bf x}_a)/2}\right)
       \nonumber \\{}
   & &  ~~~~~~~~~~~~~ ~~~~  \times \exp \left[ -\kappa
         \frac{1+\varrho }{1-\varrho } (r_b+r_a)\right],
\end{eqnarray}
with the variable
\begin{equation}
\rho \equiv e^{-2\omega L},
\label{}\end{equation}
and the parameters
\begin{eqnarray}
\nu &=&\frac{e^2}{2\omega \hbar }\frac{E}{Mc^2}=\frac{\alpha }{\sqrt{
M^2c^4/E^2-1}},\nonumber \\
\kappa& =&\frac{\mu \omega }{2\hbar }=\frac{1}{\hbar c}\sqrt{ M^2c^4-E^2}
=\frac{E}{\hbar c}\frac{\alpha}{\nu }.
\end{eqnarray}
We now
use the well-known expansion
\begin{eqnarray} \label{19.13.n5}
I_0 (z \cos (\theta /2)) = \frac{2}{z}
   \sum _{l=0}^{\infty}
         (2l+1) P_l (\cos\theta ) I_{2l+1}(z)
\end{eqnarray}
and obtain the partial wave decomposition
\begin{eqnarray} \label{19.13.n8}
({\bf x}_b|{\bf x}_a)_E
 & =  &  \frac{1}{r_br_a}\sum _{l=0}^{\infty}  (r_b|r_a)_{E,l } \frac{2l+1}
         {4\pi }
         P_l (\cos \theta )\nonumber \\
          & = & \frac{1}{r_br_a}
     \sum _{l=0}^{\infty} (r_b|r_a)_{E,l}
     \sum _{m=-l}^{l} Y_{lm} (\hat{\bf x}_b) Y^*_{lm}
          (\hat{\bf x}_a) ,
\end{eqnarray}
with the usual notation for Legendre polynomials and spherical harmonics.
The radial amplitude is, therefore,
\begin{eqnarray} \label{19.13.n9}
(r_b|r_a)_{E,l} &=&- i\frac{\hbar }{2Mc} \sqrt{ r_br_a}\frac{2M}{\hbar }
    \int_{0}^{\infty} dy \frac{1}{\sinh y}e^{2\nu y}      \\{}
   & &\times  \exp \left[ -\kappa  \coth y (r_b+r_a)\right]
      I_{2l+1}\left  (2\kappa
\sqrt{ r_br_a}\frac{1}{\sinh y}\right ).
      \nonumber
\end{eqnarray}

At this place,
 the
additional centrifugal barrier (\ref{19.13a})
is incorporated via the replacement
\begin{equation}
2l+1\rightarrow 2\tilde l+1\equiv \sqrt{ (2l+1)^2+l_{\rm extra}^2}.
\label{19.}\end{equation}
 (as in Eqs.~(8.146) and (14.237) in Ref. \cite{PI}).
The integration over $y$
yields
\begin{equation} \label{19.13.n10}
 (r_b|r_a)_{E,l} =-i\frac{\hbar }{2Mc}\frac{M}{\hbar \kappa} \frac{ \Gamma
(-\nu +\tilde l+1)}
   {(2\tilde l+1)!}  W_{\nu ,\tilde l+1/2} \left (2\kappa  r_b\right ) M_{\nu
,\tilde l+1/2}
       \left  (2\kappa r_a\right )
\end{equation}
 (compare Eq.~(9.52) in Ref. \cite{PI}).

This fixed-energy amplitude has poles in the Gamma function
whenever
$\nu -\tilde l-1=0,1,2,\dots ~$. They
determine the bound-state energies of the
Coulomb system.
Subsequent formulas
can be simplified by introducing the small
positive $l$-dependent parameter
\begin{equation}
\delta _l\equiv l- \tilde l=l+1/2-\sqrt{ (l+1/2)^2-\alpha ^2}\approx
\frac{\alpha ^2}{2l+1}+{\cal O}(\alpha ^4).
\label{19.}\end{equation}
With it, the pole positions are given by
$\nu  =\tilde n_l\equiv n-\delta _l$, with $ n=l+1, l+2, l+3, \dots~$,
and the
bound state energies become:
\begin{eqnarray}
E_{nl}&=&\pm Mc^2\left [1+\frac{\alpha ^2}{(n-\delta _l)^2}\right]^{-1/2}
\nonumber \\
&\approx &\pm Mc^2\left[ 1-\frac{\alpha ^2}{2n^2}
-\frac{\alpha ^4}{n^3}\left( \frac{1}{2l+1}-\frac{3}{8n}\right)
+{\cal O}(\alpha ^6)\right] .
\label{19.}\end{eqnarray}
Note the appearance of the plus-minus sign
as a characteristic property of
energies in relativistic quantum mechanics. A correct interpretation of the
negative energies as positive energies of antiparticles
is straightforward within quantum field theory;
it will not be discussed here.

To find the wave functions,
we approximate
near the poles $\nu \approx \tilde n _l$:
\begin{eqnarray} \label{19.13.n13}
   \Gamma (-\nu +\tilde l+1) &\approx &
-\frac{(-)^{n_r}}{n_r!} \frac{1}{\nu -\tilde n _l},\nonumber \\
\frac{1}{\nu -\tilde n_l} & \approx   &
\frac{2}{\tilde n_l}\frac{\hbar ^2\kappa ^2}{2M}
\left( \frac{E}{Mc^2}\right) ^2
       \frac{2Mc^2}{E^2-E^2_{nl}},\nonumber \\
\kappa  & \approx &\frac{E}{Mc^2}\frac{1}{a_H} \frac{1}{\tilde n_l},
\end{eqnarray}
with the {radial quantum number} $n_r=n-l-1$.
In analogy with
a corresponding nonrelativistic  equation
(Eq.~(13.203) in Ref. \cite{PI}),
the last equation can be rewritten
as
\begin{equation} \label{19.13.n11}
\kappa  =\frac{1}{\tilde a_H}\frac{1}{\nu },
\end{equation}
where
\begin{equation}
\tilde a_H\equiv a_H\frac{Mc^2}{E}
\label{}\end{equation}
denotes a modified energy-dependent
Bohr radius. Instead of being $1/\alpha\approx 137 $ times the
Compton wave length of the electron $\hbar /Mc$,
 the modified Bohr radius which sets
the length scale  of relativistic bound states
involves the energy $E$ instead of the rest energy $Mc^2$.

With the above parameters, the positive-energy
poles in the
Gamma function
can be written as
\begin{equation} \label{19.13.n14}
-i\Gamma (-\nu +\tilde l+1) \frac{M}{\hbar \kappa }
\approx \frac{(-)^{n_r}}{\tilde n _l^2 n_r!}
        \frac{1}{\tilde a_H} \left( \frac{E}{Mc^2}\right) ^2
\frac{2Mc^2i\hbar }{E^2-E^2 _{nl}}.
\end{equation}
Using this behavior and a property of
the Whittaker functions
(see Eq.~(9.73) in Ref. \cite{PI}),
we write
the
contribution of the bound states to the spectral representation
of the fixed-energy amplitude as
\begin{equation} \label{19.13.n15}
(r_b|r_a)_{E,l} = \frac{\hbar }{Mc}\sum ^{\infty} _{n=l+1} \left(
\frac{E}{Mc^2}\right) ^2
 \frac{2Mc^2 i\hbar }{E^2-E^2_{nl}}
       R_{nl}(r_b) R_{nl} (r_a)+ \dots~ .
\end{equation}
A comparison between
the
pole terms in (\ref{19.13.n10})
and (\ref{19.13.n15})
renders the radial wave functions
\begin{eqnarray} \label{19.13.n17}
R_{nl}(r)& =&\frac{1}{\tilde a_H^{1/2}\tilde  n_l }\frac{1}{(2\tilde l+1)!}
    \sqrt{ \frac{(\tilde n_l +\tilde l)!}{( n -l-1)!}}\nonumber    \\{}
   & & \times (2r/\tilde n_l \tilde a_H)^{\tilde l+1} e^{-r/\tilde n_l \tilde
a_H} M(-n +l+1,2\tilde l+2,2r/\tilde n_l \tilde a_H)
  \\{}
   & =& \frac{1}{\tilde a_H^{1/2}\tilde n_l}
   \sqrt{ \frac{(n-l-1)!}{(\tilde n+\tilde l)!}}
          e^{-r/\tilde n\tilde a_H} (2r/\tilde n_l\tilde a_H)^{\tilde l+1}
L_{\tilde n_l-l-1}^{2\tilde l+1}\nonumber
            (2r/\tilde n_l\tilde a_H).
\end{eqnarray}
The
properly normalized total wave functions are
\begin{equation} \label{19.13.n16}
\psi _{nlm}({\bf x})
=\frac{1}{r} R_{nl} (r) Y_{lm}(\hat {\bf x}).
\end{equation}

The continuous wave functions are obtained in the same way as
in the non-relativistic case
 (see Eqs.~(13.211)--(13.219) in Ref. \cite{PI}).

This concludes the solution of the path integral of
the relativistic Coulomb system.

%
%

\end{document}